\documentclass[12pt, draftclsnofoot, onecolumn]{IEEEtran}

\usepackage{enumitem}
\usepackage{subfigure,cite,epsfig,dblfloatfix}
\usepackage[cmex10]{amsmath}
\usepackage{amssymb,mathrsfs,graphicx}
\usepackage{psfrag}
\usepackage{bbm}

\let\oldbrace\{
\def\{{\oldbrace\kern0.5pt}

\let\P\relax
\DeclareMathOperator\P{\sf P}
\DeclareMathOperator*{\argmax}{arg\,max}
\DeclareMathOperator*{\argmin}{arg\,min}

\newtheorem{theorem}{Theorem}
\newtheorem{example}{Example}
\newtheorem{remark}{Remark}

\newtheorem{proposition}{Proposition}
\newtheorem{corollary}{Corollary}

\newcommand{\markov}{\mathrel\multimap\joinrel\mathrel-\mspace{-9mu}\joinrel\mathrel-}

\begin{document}

\title{Information-Theoretic Caching: \\Sequential Coding for Computing}
\author{Chien-Yi Wang, Sung Hoon Lim, and Michael Gastpar
\thanks{This work was supported in part by the European ERC Starting Grant 259530-ComCom.}
\thanks{The material in this paper was presented in part at the IEEE International Symposium on Information Theory (ISIT), Hong Kong, China, June 2015.}
\thanks{The authors are with the School of Computer and Communication Sciences, {\'E}cole Polytechnique F{\'e}d{\'e}rale de Lausanne (EPFL), Lausanne, Switzerland. Emails: \{chien-yi.wang, sung.lim, michael.gastpar\}@epfl.ch}
}

\maketitle
\begin{abstract}
Under the paradigm of caching, partial data is delivered before the actual requests of users are known. In this paper, this problem is modeled as a canonical distributed source coding problem with side information, where the side information represents the users' requests. For the single-user case, a single-letter characterization of the optimal rate region is established, and for several important special cases, closed-form solutions are given, including the scenario of uniformly distributed user requests. In this case, it is shown that the optimal caching strategy is closely related to total correlation and Wyner's common information. Using the insight gained from the single-user case, three two-user scenarios admitting single-letter characterization are considered, which draw connections to existing source coding problems in the literature: the Gray--Wyner system and distributed successive refinement. Finally, the model studied by Maddah-Ali and Niesen is rephrased to make a comparison with the considered information-theoretic model. Although the two caching models have a similar behavior for the single-user case, it is shown through a two-user example that the two caching models behave differently in general. 
\end{abstract}


%-----------------------------------------------------------------------------

\section{Introduction}
Consider a sports event filmed simultaneously by many cameras. After the game, a sports aficionado would like to watch a customized video sequence on his mobile device that shows him, in every moment, the best angle on his favorite player in the game. Of course, he would like that video as soon as possible. To meet such demand, the provider could choose to use the paradigm of caching: even before knowing the precise camera angles of interest, cleverly coded partial data is delivered to the end user device. If all goes well, that partial data is at least partially useful, and hence, at delivery time, a much smaller amount of data needs to be downloaded, leading to faster (and possibly cheaper) service. To make matters even more interesting, there might be several users in the same mobile cell with the same wish, except that they probably have different favorite players. Now, the caching technique could be employed at the base station of the mobile cell, and the goal is to design cache contents such that at delivery time, almost all users experience a faster download speed. 

In the present paper, we model this situation in a canonical information-theoretic fashion. We model the data as a long sequence $X_1, X_2, X_3, \cdots,$ where the subscript may represent the time index. That is, in the above example, $X_i$ would represent the full collection of video images acquired at time $i.$ Furthermore, in our model, the user defines a separate request for each time instant $i.$ Hence, the user's requests are also modeled as a sequence $Y_1, Y_2, Y_3, \cdots,$ whose length we assume to be identical to the length of the data sequence. That is, in the above example, $Y_i$ represents the user's desired camera angle at time $i.$ There are two encoders: The cache encoder and the update encoder. The cache encoder only gets to observe the data sequence, and encodes it using an average rate of $R_{\sf c}$ bits per symbol. The update (or data delivery) encoder gets to observe both the data sequence and the request sequence, and encodes them jointly using an average rate of $R_{\sf u}$ bits per symbol. At the decoding end, to model the user's desired data, we consider a per-letter function $g(X_i, Y_i)$ which needs to be recovered losslessly for all $i.$ For example, we may think of $X_i$ as being a vector of a certain length, and of $Y_i$ as the index of the component of interest to the user at time $i.$ Then, $g(X_i, Y_i)$ would simply return the component indexed by $Y_i$ from the vector $X_i.$ The goal of this paper is to characterize the set of those rate pairs $(R_{\sf c}, R_{\sf u})$ that are sufficient to enable the user to attain his/her goal of perfectly recovering the entire sequence $g(X_1, Y_1), g(X_2, Y_2), g(X_3, Y_3), \cdots.$ 

When there are multiple users, normally different end users have distinct requests and functions, denoted by $Y_i^{(\ell)}$ and $g_\ell(\cdot, \cdot)$, respectively, for end user $\ell$. Moreover, different end users may share caches and updates. However, in this work we make the simplification that each user's request is known to all users.\footnote{In practice, this can be realized by the server broadcasting requests to all users. Comparing with the desired data itself, usually the amount of information contained in the requests is much less and it is even more so when the request distribution is far from uniform. Then, the penalty due to the overhead of revealing requests will be acceptable for some applications.} That is, denoting by $Y_i = \{Y_i^{(\ell)}\}$ the collection of requests, we assume that $Y_i$ is globally known except to the cache encoder. Thus, we denote $f_\ell(X_i,Y_i) = g_\ell(X_i,Y_i^{(\ell)})$ and focus on the functions $\{f_\ell(\cdot,\cdot)\}$ hereafter.

An important inspiration for the work presented here are the pioneering papers of Maddah-Ali and Niesen \cite{Maddah-Ali:14, Niesen:14}. Their work emphasizes the case when there is a large number of users and develops clever caching strategies that centrally leverage a particular multi-user advantage: Cache contents is designed in such a way that each update (or delivery) message is simultaneously useful for as many users as possible. Our present work places the emphasis on the statistical properties of the data and requests, exploiting these features in a standard information-theoretic fashion by coding over multiple instances.

On the modeling side, one could relate the Maddah-Ali--Niesen model to our model in the following manner: Consider a single data $X$ and a single request $Y$. This $X$ is composed of $N$ files, each containing $F$ bits, and the $Y$ designates the index of the desired file. Then, the considered information-theoretic model corresponds to coding over multiple instances of $(X,Y)$, assuming $F$ is a fixed constant. On the other hand, the Maddah-Ali--Niesen model corresponds to coding over a single instance of $(X,Y)$ but the file size $F$ can be arbitrarily large. In this paper, the Maddah-Ali--Niesen model will also be referred to as the ``static request" model. Ample results are available for the static request model at this point: the worst-case analysis \cite{Maddah-Ali:14}, the average-case analysis \cite{Niesen:14}, decentralized \cite{Maddah-Ali:15}, delay-sensitive \cite{Niesen:14d}, online \cite{Pedarsani:14}, multiple layers \cite{Karamchandani14}, request of multiple items \cite{Ji14m}, secure delivery \cite{Sengupta:15}, wireless networks \cite{Ji14b, Hachem14}, etc. In addition, some improved order-optimal results for the average case can be found in \cite{Ji14z, Zhang:15}.

The Maddah-Ali--Niesen model fits well with applications in which the users' requests remain fixed over the entire time period of interest, e.g., on-demand video streaming of a movie from a given database. By contrast, our model may be an interesting fit for applications in which the users' requests change over time, such as the multi-view video system example in the beginning. Furthermore, our model fits well with sensor network applications. In many cases, only the sensor data (modeled as $X$) with certain properties (modeled as $Y$) are of interest and the desired properties may vary over a timescale of minutes, hours, or even days. In terms of coding strategies, we also take a different approach from \cite{Maddah-Ali:14, Niesen:14} and the follow-up works, in which the core schemes are based on linear network coding. By contrast, by formulating the caching problem into a multi-terminal source coding problem, our main tools are standard information-theoretic arguments, including joint typicality encoding/decoding, superposition coding, and binning. 

\begin{figure}[t!]
\begin{center}
\includegraphics[scale=0.75]{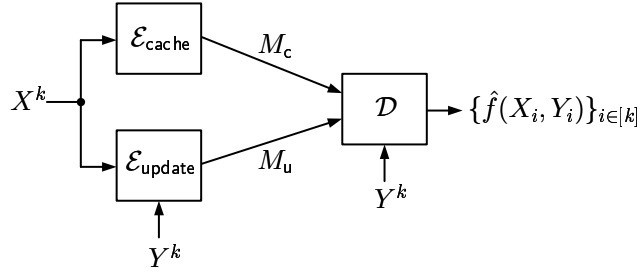}
\end{center}
\vspace{-0.25in}
\caption{The single-user case.}
\label{fig:single}
\vspace{-0.1in}
\end{figure}

\subsection{Related Works}
The structure of many source coding problems studied in the literature can be captured by our formulation. Let us start with the single-user case, as shown in Figure \ref{fig:single}. Denote by $R_{\sf c}$ and $R_{\sf u}$ as the rates of the cache and the update, respectively. Depending on the availability of the cache and the update, each configuration can be seen as a special case or a straightforward extension of 
\begin{enumerate}
\item $(0,R_{\sf u})$: lossless source coding with side information \cite{Slepian:73}; 
\item $(R_{\sf c},0)$: lossy source coding with side information \cite{Wyner:76} or lossless coding for computing with side information \cite{Orlitsky:01};
\item $(R_{\sf c},R_{\sf u})$: lossless source coding with a helper \cite{Wyner:75b}, \cite{Ahlswede:75}.
\end{enumerate}

As for the two-user case, two classes of source coding problems are related to our problem setup: the Gray--Wyner system and the problem of successive refinement.

\begin{figure}[t!]
\begin{center}
\includegraphics[scale=0.7]{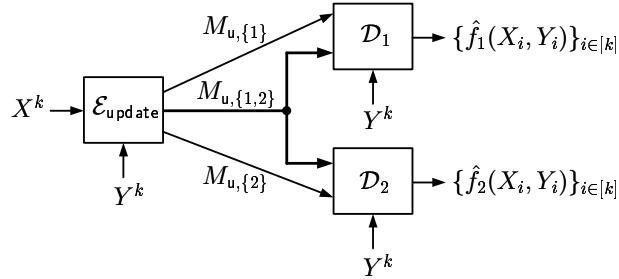}
\end{center}
\vspace{-0.25in}
\caption{The Gray--Wyner system in the considered setup.}
\label{fig:GW}
\vspace{-0.1in}
\end{figure}

In the Gray--Wyner system, each decoder receives a private message and a common message. A system diagram of the considered setup is shown in Figure \ref{fig:GW}. The optimal rate--distortion region was characterized in \cite{Gray:74}. The extension to include distinct side information at the decoders was studied in \cite{Timo:08} and \cite[Section V]{Shayevitz:13}.

\begin{figure}[t!]
\begin{center}
\includegraphics[scale=0.7]{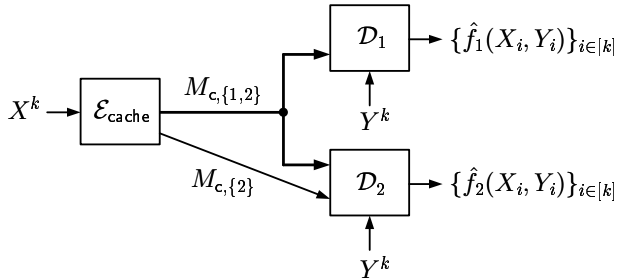}
\end{center}
\vspace{-0.25in}
\caption{The problem of successive refinement in the considered setup.}
\label{fig:SuccRefine}
\vspace{-0.1in}
\end{figure}

In the problem of successive refinement \cite{Koshelev:85, Equitz:91}, one of the decoders has no private message, i.e., all of its received messages are also available at the other decoder. A system diagram in the considered setup is shown in Figure \ref{fig:SuccRefine}. The optimal rate--distortion region was characterized in  \cite{Rimoldi:94}. The extension to include distinct side information at the decoders was studied in \cite{Steinberg:04, Tian:08, Akyol:14b}. The problem of successive refinement can also be extended to multiple sources \cite{CYWang:14, Akyol:14a}. One special case of the multi-source extension is the problem of sequential coding of correlated sources \cite{Viswanathan:00}.

\subsection{Summary of Results}
In this work we make some progress for the single-user case and extend the results to some two-user scenarios. For the single-user case, the results are summarized as follows: 
\begin{itemize}[leftmargin=*]
\item Theorem \ref{thm:single} provides a single-letter characterization of the optimal rate region.
\item Propositions \ref{prop:ind} and \ref{prop:nest} give the exact optimal rate regions for the cases of independent components and nested components, respectively, and confirm that some intuitive caching strategies are indeed optimal.
\item Proposition \ref{prop:uni} shows that if the components are uniformly requested, then the optimal caching strategy is to cache a description of the data that minimizes the conditional total correlation.
\end{itemize}
For the two-user case, we find single-letter expressions for the following scenarios: 
\begin{itemize}[leftmargin=*]
\item the private-update-aided Gray--Wyner system (Theorem \ref{thm:u2c1});
\item the common-cache-aided Gray--Wyner system (Theorem \ref{thm:u2c2});
\item the problem of sequential successive refinement (Theorem \ref{thm:u2c3}).
\end{itemize}

In addition, we show that the information-theoretic model and the average-case formulation of the Maddah-Ali--Niesen model have the same scaling behavior for the single-user case. However, through a two-user example, we show that in general the two caching models behave differently and coding over multiple blocks can be beneficial.

The paper is organized as follows. In Section \ref{sec:prob_state}, we provide the problem formulation for the general two-user case. 
Section \ref{sec:single} is devoted to the single-user case. In Section \ref{sec:two_user}, we consider several two-user scenarios which are extensions of the Gray--Wyner system and the problem of successive refinement. In Section \ref{sec:static} we go back to the single-user case and discuss the static request model. Finally, we conclude in Section \ref{sec:conclude}. The lengthy proofs are deferred to appendices.

\subsection{Notations}
Random variables and their realizations are represented by uppercase letters (e.g., $X$) and lowercase letters (e.g., $x$), respectively. We use calligraphic symbols (e.g., $\mathcal{X}$) to denote sets. Denote by $|\cdot|$ the cardinality of a set. 
We denote $[a] := \{1,2,\cdots, \lfloor a \rfloor\}$ for all $a\ge 1$ and $\mathcal{A}\backslash\mathcal{B} :=\{x\in\mathcal{A}|x\notin\mathcal{B}\}$. We denote $X^k := (X_1,X_2,\cdots,X_k)$ and $X^{[k]\backslash\{i\}} := (X_1,\cdots,X_{i-1},X_{i+1},\cdots,X_k)$. Throughout the paper, all logarithms are to base two. Let $h_{\sf b}(p):=-p\log(p)-(1-p)\log(1-p)$ for $p\in[0,1]$ and $0\log(0) :=0$ by convention. We denote $(x)^+ := \max\{x,0\}$ and follow the $\epsilon$--$\delta$ notation in \cite{ElGamal:11}. 

\section{Problem statement} \label{sec:prob_state}
For convenience, we provide a problem statement for the general two-user case and then restrict attention to four special cases of interest. Consider a discrete memoryless source (DMS) $\langle X,Y\rangle$ with finite alphabet $\mathcal{X}\times\mathcal{Y}$ and joint probability mass function (pmf) $p_{X,Y}$. The DMS $\langle X,Y\rangle$ generates an independent and identically distributed (i.i.d.) random process $((X_i,Y_i):i\in\mathbb{Z}^+)$. 

Denote by $k$ a positive integer. There are two encoding terminals and two decoding terminals. The {\em cache} encoder observes the source sequence $X^k$, the {\em update} encoder observes the source sequences $(X^k,Y^k)$, and both decoders observe $Y^k$. Decoder $\ell \in \{1,2\}$ wishes to recover an element-wise function $f_\ell(x,y)$ losslessly. The cache encoder generates three messages $M_{{\sf c},\{1\}}$, $M_{{\sf c},\{2\}}$, and $M_{{\sf c},\{1,2\}}$ of rate $R_{{\sf c},\{1\}}$, $R_{{\sf c},\{2\}}$, and $R_{{\sf c},\{1,2\}}$, respectively. Similarly, the update encoder generates three messages $M_{{\sf u},\{1\}}$, $M_{{\sf u},\{2\}}$, and $M_{{\sf u},\{1,2\}}$ of rate $R_{{\sf u},\{1\}}$, $R_{{\sf u},\{2\}}$, and $R_{{\sf u},\{1,2\}}$, respectively. We note that some of these rates may be zero. Then, Decoder $\ell\in\{1,2\}$ receives the set of messages $(M_{{\sf c},\{\ell\}}, M_{{\sf c},\{1,2\}}, M_{{\sf u},\{\ell\}}, M_{{\sf u},\{1,2\}})$. The messages $M_{{\sf c},\{\ell\}}$ and $M_{{\sf u},\{\ell\}}$ are called {\it private cache content} and {\it private update content} of User $\ell\in\{1,2\}$, respectively. Besides, the messages $M_{{\sf c},\{1,2\}}$ and $M_{{\sf u},\{1,2\}}$ are called {\it common cache content} and {\it common update content}, respectively.

For convenience, we denote 
\begin{IEEEeqnarray*}{rCl}
\mathbf{R} &:=& (R_{{\sf c},\{1\}},R_{{\sf c},\{2\}},R_{{\sf c},\{1,2\}},R_{{\sf u},\{1\}},R_{{\sf u},\{2\}},R_{{\sf u},\{1,2\}}), \\
2^{k\mathbf{R}} &:=& (2^{kR_{{\sf c},\{1\}}},2^{kR_{{\sf c},\{2\}}},2^{kR_{{\sf c},\{1,2\}}},2^{kR_{{\sf u},\{1\}}},2^{kR_{{\sf u},\{2\}}},2^{kR_{{\sf u},\{1,2\}}}).
\end{IEEEeqnarray*}
Then, a $(2^{k\mathbf{R}},k)$ distributed multiple description code consists of ($\mathcal{A}\in \{\{1\},\{2\},\{1,2\}\}$)
\begin{itemize}
\item two encoders, where the cache encoder assigns three indices $m_{{\sf c},\mathcal{A}}(x^k)\in[2^{kR_{{\sf c},\mathcal{A}}}]$ to each sequence $x^k\in\mathcal{X}^k$ and the update encoder assigns three indices $m_{{\sf u},\mathcal{A}}(x^k,y^k)\in[2^{kR_{{\sf u},\mathcal{A}}}]$ to each pair of sequences $(x^k,y^k)\in\mathcal{X}^k\times\mathcal{Y}^k$;  
\item two decoders, where Decoder $\ell\in \{1,2\}$ assigns an estimate $\hat{s}_\ell^k$ to each tuple \\$(m_{{\sf c},\{\ell\}}, m_{{\sf c},\{1,2\}}, m_{{\sf u},\{\ell\}}, m_{{\sf u},\{1,2\}},y^k)$.
\end{itemize}

A rate tuple $\mathbf{R}$ is said to be achievable if there exists a sequence of $(2^{k\mathbf{R}},k)$ codes with 
\begin{IEEEeqnarray*}{rCl}
\lim_{k\to\infty} \P\left(\bigcup_{\ell\in\{1,2\}}\bigcup_{i\in[k]}\left\{\hat{S}_{\ell,i}\neq f_\ell(X_i,Y_i)\right\}\right) &=& 0.
\end{IEEEeqnarray*}
Note that the probability is taken with respect to both $X^k$ and $Y^k$.
The optimal rate region $\mathcal{R}_{\sf full}^\star$ is the closure of the set of achievable rate tuples.

In this paper, we only consider the following projections of $\mathcal{R}_{\sf full}^\star$\footnote{More results on the general two-user case can be found in \cite[Chapter $4$]{CYWang:15}, including a full achievable rate region.}, in which some of the rate components are set to zero: 
\begin{enumerate}[leftmargin=*]
\item the single-user case 
\begin{IEEEeqnarray*}{rCl}
\mathcal{R}^\star &:=& \{(R_{{\sf c},\{1\}},R_{{\sf u},\{1\}}): (R_{{\sf c},\{1\}},0,0,R_{{\sf u},\{1\}},0,0)\in\mathcal{R}_{\sf full}^\star\};
\end{IEEEeqnarray*}
\item private-update-aided Gray--Wyner system 
\begin{IEEEeqnarray*}{rCl}
\mathcal{R}^\star_{\sf puGW} &:=& \{(R_{{\sf c},\{1\}},R_{{\sf c},\{2\}},R_{{\sf c},\{1,2\}},R_{{\sf u},\{1\}},R_{{\sf u},\{2\}}): \\
&& \hspace{1cm} (R_{{\sf c},\{1\}},R_{{\sf c},\{2\}},R_{{\sf c},\{1,2\}},R_{{\sf u},\{1\}},R_{{\sf u},\{2\}},0)\in\mathcal{R}_{\sf full}^\star\};
\end{IEEEeqnarray*}
\item common-cache-aided Gray--Wyner system
\begin{IEEEeqnarray*}{rCl}
\mathcal{R}^\star_{\sf ccGW} &:=& \{(R_{{\sf c},\{1,2\}},R_{{\sf u},\{1\}},R_{{\sf u},\{2\}},R_{{\sf u},\{1,2\}}): \\
&& \hspace{1cm} (0,0,R_{{\sf c},\{1,2\}},R_{{\sf u},\{1\}},R_{{\sf u},\{2\}},R_{{\sf u},\{1,2\}})\in\mathcal{R}_{\sf full}^\star\};
\end{IEEEeqnarray*}
\item sequential successive refinement
\begin{IEEEeqnarray*}{rCl}
\mathcal{R}^\star_{\sf SSR} &:=& \{(R_{{\sf c},\{2\}},R_{{\sf c},\{1,2\}},R_{{\sf u},\{2\}},R_{{\sf u},\{1,2\}}): \\
&& \hspace{1cm} (0,R_{{\sf c},\{2\}},R_{{\sf c},\{1,2\}},0,R_{{\sf u},\{2\}},R_{{\sf u},\{1,2\}})\in\mathcal{R}_{\sf full}^\star\}.
\end{IEEEeqnarray*}
\end{enumerate} 
With an abuse of notation, members of different projections are simply denoted by $\mathbf{R}$ and its dimension will be clear from context. For the single-user case, all the decoder indices are dropped, e.g., $f_1(x,y)$ is simply denoted by $f(x,y)$.

\section{The Single-User Case} \label{sec:single}
In this section, we present the main results for the single-user case. This setup is a special case of the problem of lossy source coding with a helper (and decoder side information), which remains open in general. We establish the optimal rate region of the considered setup in the following theorem. The proof is deferred to Appendix \ref{sec:proof_single}.
\begin{theorem}
\label{thm:single}
Consider the single-user caching problem. The optimal rate region $\mathcal{R}^\star$ is the set of rate pairs $(R_{\sf c},R_{\sf u})$ such that 
\begin{IEEEeqnarray}{rCl}
\label{eq:single1}
R_{\sf c} &\ge& I(X;V|Y), \\
\label{eq:single2}
R_{\sf u} &\ge& H(f(X,Y)|V,Y),
\end{IEEEeqnarray} 
for some conditional pmf $p_{V|X}$ with $|\mathcal{V}| \le |\mathcal{X}|+1$.
\end{theorem}

Due to the fact that the update encoder is more informative than the cache encoder, we have the following corollary from Theorem \ref{thm:single}. We denote $R_{\sf u}^\star := \min \{R_{\sf u}: (0,R_{\sf u})\in\mathcal{R}^\star\}$ and $R_{\sf c}^\star := \min \{R_{\sf c}: (R_{\sf c},0)\in\mathcal{R}^\star\}$. The proof is deferred to Appendix \ref{sec:proof_property}.
\begin{corollary} \label{col:single}
Consider the single-user caching problem. All the following statements hold:
\begin{enumerate}[leftmargin=*]
\item $R_{\sf u}^\star = H(f(X,Y)|Y)$.
\item $R_{\sf c}^\star = \min I(X;V|Y)$, where the minimum is over all conditional pmfs $p_{V|X}$ satisfying $H(f(X,Y)|V,Y)=0$.
\item $R_{\sf c}+R_{\sf u}\ge H(f(X,Y)|Y)$ for all $(R_{\sf c},R_{\sf u}) \in \mathcal{R}^\star$.
\item $R_{\sf u}^\star \le R_{\sf c}^\star$.
\item If $(R_{\sf c},R_{\sf u}) \in \mathcal{R}^\star$, then for all $a\in[0,R_{\sf c}]$, $(R_{\sf c}-a,R_{\sf u}+a)\in \mathcal{R}^\star$.
\item If $(R_{\sf c},R_{\sf u})$ is an extreme point of $\mathcal{R}^\star$ with $R_{\sf c}>0$, then for all $a> 0$, $(R_{\sf c}+a,R_{\sf u}-a)\notin \mathcal{R}^\star$.
\end{enumerate}
\end{corollary}
Note that Statement $2$ of Corollary \ref{col:single} recovers the result of lossless coding for computing with side information \cite{Orlitsky:01}. Statements $5$ and $6$ point out in which direction we can move a partial rate such that the resulting rate pair still resides in $\mathcal{R}^\star$.

In general, the optimal sum rate can only be achieved with $(R_{\sf c},R_{\sf u}) = (0,H(f(X,Y)|Y))$, i.e., the update encoder does all the work. Namely, in general there is a penalty when the request $Y^k$ is not known at the encoder.
Nevertheless, for the class of partially invertible functions, one can arbitrarily distribute the work load without compromising the sum rate. 
\begin{corollary} \label{col:SI_useless}
If the function $f$ is partially invertible, i.e., $H(X|f(X,Y),Y)=0$, then $R_{\sf c}^\star = R_{\sf u}^\star = H(X|Y)$. 
\end{corollary}
\begin{IEEEproof}
First, Corollary \ref{col:single} says that $H(f(X,Y)|Y) = R_{\sf u}^\star \le R_{\sf c}^\star \le H(X|Y)$, where the last inequality follows by setting $V=X$ in \eqref{eq:single1}. Then, the corollary follows immediately by noting that  
\begin{IEEEeqnarray*}{rCCCl}
H(f(X,Y)|Y) &=& H(f(X,Y),X|Y) &=& H(X|Y), 
\end{IEEEeqnarray*}
where the first equality follows since $H(X|f(X,Y),Y)=0$, by assumption.
\end{IEEEproof}
In other words, Corollary \ref{col:SI_useless} says that for partially invertible functions, e.g., arithmetic sum and modulo sum, the side information $Y$ at the update encoder is useless in lowering the compression rate and thus in this case the cache encoder is as powerful as the update encoder. More generally, it can be shown that $R_{\sf c}^\star = R_{\sf u}^\star$ if and only if there exists a conditional pmf $p_{V|X}$ such that  
\begin{enumerate}
\item $H(V|f(X,Y),Y) = H(V|X,Y)$, and 
\item $H(f(X,Y)|V,Y) = 0$.
\end{enumerate}

For most single-user caching problems, it is challenging to find a closed-form expression for the optimal rate region $\mathcal{R}^\star$. In words, we do not know the optimal caching strategy in general. In the following, we consider three cases where $X$ and $Y$ are independent, which implies that $I(X;V|Y) = I(X;V)$. For the first two cases, we are able to show that some intuitive caching strategies are indeed optimal. In the last case, we provide some guidance for the optimal caching strategy. Without loss of generality, we assume that $\mathcal{Y}=[N]$. Besides, we will find it convenient to denote $x^{(y)} := f(x,y)$. 

\begin{remark}
If $X$ and $Y$ are independent, then the optimal conditional pmf $p^\star_{V|X}$ for a fixed cache rate $R_{\sf c}$ can be found by solving the following constrained optimization problem 
\begin{IEEEeqnarray*}{rCl}
\text{maximize} &\hspace{1cm}& I(f(X,Y),Y;V) \\
\text{subject to} &\hspace{1cm}& I(X;V) \le R_{\sf c}
\end{IEEEeqnarray*}
over all conditional pmf $p_{V|X}$ with $|\mathcal{V}|\le|\mathcal{X}|+1$. To see this, we observe that 
\begin{IEEEeqnarray*}{rCl}
\argmin_{p_{V|X}} H(f(X,Y)|V,Y) &=& \argmax_{p_{V|X}} H(f(X,Y)|Y) - H(f(X,Y)|V,Y) \\
&=& \argmax_{p_{V|X}} I(f(X,Y);V|Y) \\
&\overset{(a)}{=}& \argmax_{p_{V|X}} I(f(X,Y),Y;V), 
\end{IEEEeqnarray*}
where $(a)$ follows since $X$ and $Y$ are independent, by assumption, and $V\markov X \markov Y$ form a Markov chain.
Thus, caching has an {\em information bottleneck} interpretation \cite{Tishby:99} (see also \cite{Witsenhausen:75}). Given a fixed-size cache as bottleneck, we aim to provide the most relevant information of the desired function $f(X,Y)$. The existing algorithms developed for the information bottleneck problem are applicable to numerically approximate the optimal rate region $\mathcal{R}^\star$. 
\hfill$\lozenge$\end{remark}

\subsection{Independent Source Components}
In this subsection, we consider the case where the source components $X^{(1)},\cdots,X^{(N)}$ are independent. Note that we used the short-hand notation $X^{(n)} = f(X,n)$. Without loss of generality, we assume that $p_Y(1)\ge p_Y(2) \ge \cdots \ge p_Y(N)$. Then, we have the following proposition.
\begin{proposition}\label{prop:ind}
If $X$ and $Y$ are independent and the source components $X^{(1)},\cdots,X^{(N)}$ are independent, then the optimal rate region $\mathcal{R}^\star$ is the set of rate pairs $(R_{\sf c},R_{\sf u})$ such that 
\begin{IEEEeqnarray*}{rCl}
R_{\sf c} &\ge& r, \\
\label{eq:ind2}
R_{\sf u} &\ge& \sum_{n=1}^N (p_Y(n)-p_Y(n+1)) \left(\sum_{j=1}^nH(X^{(j)})-r\right)^+, 
\IEEEyesnumber
\end{IEEEeqnarray*}
for some $r\ge0$, where $p_Y(N+1)=0$.  
\end{proposition}
When relating to the motivating example in Introduction, Proposition \ref{prop:ind} indicates that the best caching strategy for independent views is to cache the most popular ones, no matter how different the video qualities are (see also \cite{Niesen:14} and the references therein). More generally, when the user wants to retrieve multiple views at the same time, caching the most popular ones remains optimal. The details can be found in \cite[Theorem 6]{WLG_ITA:16}.

\begin{IEEEproof}
Here we prove the achievability part. The converse part is deferred to Appendix \ref{sec:conv_ind}. Note that \eqref{eq:ind2} is equivalent to saying that 
\begin{enumerate}
\item if $r \ge \sum_{n=1}^NH(X^{(n)})$, then $R_{\sf u} \ge 0$, and 
\item if $\sum_{j=1}^{n-1} H(X^{(j)}) \le r < \sum_{j=1}^n H(X^{(j)})$ for some $n\in[N]$, then  
\begin{IEEEeqnarray*}{rCl}
R_{\sf u} &\ge& p_Y(n)\left(\sum_{j=1}^nH(X^{(j)})-r\right) + \sum_{j=n+1}^N p_Y(j) H(X^{(j)}).
\end{IEEEeqnarray*}
\end{enumerate}
Therefore, for all $n\in[0:N]$, setting $V=(X^{(1)},\cdots,X^{(n)})$ in \eqref{eq:single1} and \eqref{eq:single2} shows that the rate pair 
\begin{IEEEeqnarray*}{rCl}
(R_{\sf c},R_{\sf u}) &=& \left(\sum_{j=1}^nH(X^{(j)}),\sum_{j=n+1}^Np_Y(j)H(X^{(j)})\right)  
\end{IEEEeqnarray*}
is achievable, which corresponds to a corner point of the region described by $R_{\sf c}\ge r$ and \eqref{eq:ind2}. Since the rest of points on the boundary can be achieved by memory sharing, the proposition is established.
\end{IEEEproof}

\subsection{Nested Source Components}
Again using the shorthand notation $X^{(n)} = f(X,n)$, in this subsection we assume that $H(X^{(n)}|X^{(n+1)}) = 0$ for all $n\in[N-1]$. Then, we have the following proposition.
\begin{proposition} \label{prop:nest}
If $X$ and $Y$ are independent and $H(X^{(n)}|X^{(n+1)}) = 0$ for all $n\in[N-1]$, then the optimal rate region $\mathcal{R}^\star$ is the set of rate pairs $(R_{\sf c},R_{\sf u})$ such that 
\begin{IEEEeqnarray*}{rCl}
R_{\sf c} &\ge& r, \\
\label{eq:nest2}
R_{\sf u} &\ge& \sum_{n=1}^N p_Y(n)\left(H(X^{(n)})-r\right)^+, \IEEEyesnumber
\end{IEEEeqnarray*}
for some $r\ge0$.  
\end{proposition}
If we think of $X^{(1)},\cdots,X^{(N)}$ as representations of the same view but with different levels of quality, then Proposition $2$ indicates that the best caching strategy is to cache the finest version that still fits into the cache. 

\begin{IEEEproof}
Here we prove the achievability part. The converse part is deferred to Appendix \ref{sec:conv_nest}. Note that \eqref{eq:nest2} is equivalent to saying that 
\begin{enumerate}
\item if $r \ge H(X^{(N)})$, then $R_{\sf u} \ge 0$, and 
\item if $H(X^{(j-1)}) \le r < H(X^{(j)})$ for some $j\in[N]$, where $H(X^{(0)}):=0$, then  
\begin{IEEEeqnarray*}{rCl}
R_{\sf u} &\ge& \sum_{n=j}^N p_Y(n)\left(H(X^{(n)})-r\right).
\end{IEEEeqnarray*}
\end{enumerate}
Therefore, for all $n\in[0:N]$, substituting $V=X^{(n)}$ in \eqref{eq:single1} and \eqref{eq:single2} shows that the rate pair 
\begin{IEEEeqnarray*}{rCl}
(R_{\sf c},R_{\sf u}) &=& \left(H(X^{(n)}),\sum_{j=n+1}^Np_Y(j)H(X^{(j)}|X^{(n)})\right) 
\end{IEEEeqnarray*}
is achievable, which corresponds to a corner point of the region described by $R_{\sf c}\ge r$ and \eqref{eq:nest2}. Since the rest of points on the boundary can be achieved by memory sharing, the proposition is established.
\end{IEEEproof}

\subsection{Arbitrarily Correlated Components with Uniform Requests}
Here we assume that the request is uniformly distributed, i.e., $p_Y(n)=\frac{1}{N}$ for all $n\in[N]$, but $X^{(1)},\cdots,X^{(N)}$ can be arbitrarily correlated. Recall that $X^{(n)} = f(X,n)$. Although we cannot give a closed-form expression of the optimal rate region, we provide a necessary and sufficient condition on the auxiliary random variable which characterizes the boundary of the optimal rate region. The proof is deferred to Appendix \ref{sec:proof_uni}.

\begin{proposition} \label{prop:uni}
If $X$ and $Y$ are independent and $p_Y(n)=\frac{1}{N}$ for all $n\in[N]$, then all points $(R_{\sf c},R_{\sf u})$ on the boundary of the optimal rate region $\mathcal{R}^\star$ can be expressed as 
\begin{IEEEeqnarray*}{rCl}
R_{\sf c} &=& r, \\
R_{\sf u} &=& \frac{1}{N}\left(H(\overline{X})-r+\min_{\substack{ p_{V|\overline{X}} \\  \text{s.t. } I(\overline{X};V) = r}}\Gamma(\overline{X}|V)\right), 
\end{IEEEeqnarray*}
for some $r \in [0,H(\overline{X})]$, where $\overline{X} := (X^{(1)},X^{(2)},\cdots,X^{(N)})$ and 
\begin{IEEEeqnarray*}{rCl}
\Gamma(\overline{X}|V) &:=& \left[\sum_{n=1}^N H(X^{(n)}|V)\right] - H(X^{(1)},\cdots,X^{(N)}|V).
\end{IEEEeqnarray*}
\end{proposition}
If $N=2$, we have 
\begin{IEEEeqnarray*}{rCl}
\Gamma(\overline{X}|V) &=& I(X^{(1)};X^{(2)}|V), 
\end{IEEEeqnarray*}
so the term $\Gamma(X^{(1)},\cdots,X^{(N)}|V)$ can be interpreted as a generalization of conditional mutual information. 
In fact, the term $\Gamma(X^{(1)},\cdots,X^{(N)}) = \left[\sum_{n=1}^N H(X^{(n)})\right] - H(X^{(1)},\cdots,X^{(N)})$ was first studied by Watanabe \cite{Watanabe:60} and given the name {\it total correlation}. Following this convention, we refer to $\Gamma(X^{(1)},\cdots,X^{(N)}|V)$ as {\it conditional total correlation}. Proposition \ref{prop:uni} indicates that an optimal caching strategy is to cache a description of the data that minimizes the conditional total correlation. 

When the cache rate is large enough, there exists a conditional pmf $p_{V|X}$ such that the conditional total correlation is zero and thus we have the following corollary. 
\begin{corollary} \label{col:uni}
The boundary of the region $\{(R_{\sf c},R_{\sf u})\in\mathcal{R}^\star | R_{\sf crit} \le R_{\sf c}\le H(\overline{X})\}$ is a straight line $R_{\sf c}+NR_{\sf u} = H(\overline{X})$, where 
\begin{IEEEeqnarray*}{rCl}
R_{\sf crit} &=& \min_{\substack{p_{V|\overline{X}} \\  \text{s.t. } \Gamma(\overline{X}|V) = 0}} I(\overline{X};V).
\end{IEEEeqnarray*}
\end{corollary}
Note that when $N=2$, $R_{\sf crit}$ is Wyner's common information \cite{Wyner:75a}. 

\begin{figure}[t!]
\centering
\includegraphics[scale=1]{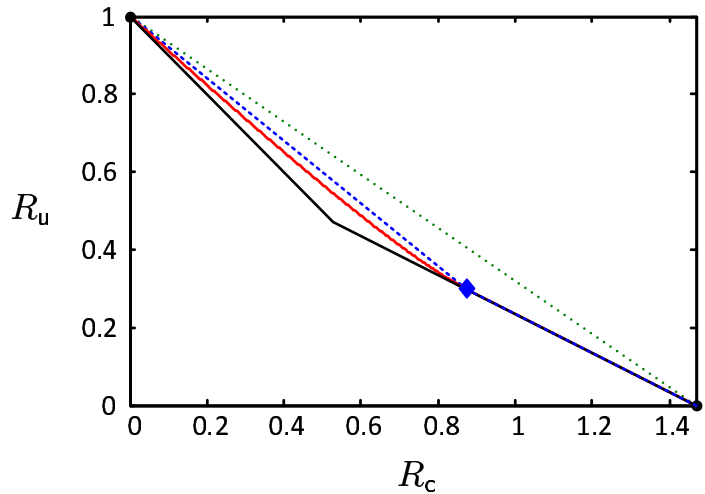}
\vspace{-0.05in}
\caption{Inner bounds and an outer bound for Example \ref{ex:sz}. Here $q=0.1$.}
\label{fig:ex_single}
\end{figure}

Finally, let us consider an example which covers all three mentioned cases.
\begin{example} \label{ex:sz}
Fix $q\in[0,\frac{1}{2}]$. Consider $Y\sim$ Uniform$(\{1,2\})$ and $X = (X^{(1)},X^{(2)})$, where $\langle X^{(1)},X^{(2)}\rangle$ is a DSBS($q$). Assume that $X$ and $Y$ are independent. We first consider two extreme cases. 
\begin{enumerate}[leftmargin=*]
\item If $q = 1/2$, then the two components are independent and $\mathcal{R}^\star = \{(R_{\sf c},R_{\sf u}) : R_{\sf c}\ge 0, R_{\sf u}\ge 0, R_{\sf c}+2R_{\sf u}\ge 2\}$.
\item If $q=0$, then the two components are nested and $\mathcal{R}^\star = \{(R_{\sf c},R_{\sf u}) : R_{\sf c}\ge 0, R_{\sf u}\ge 0, R_{\sf c}+R_{\sf u}\ge 1\}$.
\end{enumerate}

Now consider $0<q<\frac{1}{2}$. Wyner's common information of $(X^{(1)},X^{(2)})$ is known as \cite{Wyner:75a}
\begin{IEEEeqnarray*}{rCl}
R_{\sf crit} &=& 1 + h_{\sf b}(q) - 2h_{\sf b}(q'), 
\end{IEEEeqnarray*}
where $q' = \frac{1}{2}(1-\sqrt{1-2q})$. Thus, from Corollary \ref{col:uni}, we have 
\begin{IEEEeqnarray*}{rCl}
\min \{R_{\sf u} : R_{\sf c}\ge R_{\sf crit}, (R_{\sf c},R_{\sf u})\in \mathcal{R}^\star\} 
&=& \frac{1}{2}(1+h_{\sf b}(q)-R_{\sf c}). 
\end{IEEEeqnarray*}
Note that $R_{\sf u} \ge \frac{1}{2}(1+h_{\sf b}(q)-R_{\sf c})^+ $ is also a valid lower bound for all $R_{\sf c}\ge 0$. Besides, from Statement $3$ of Corollary \ref{col:single} we have $R_{\sf c} + R_{\sf u} \ge 1$.

As for the case $0<q<\frac{1}{2}$ and $R_{\sf c} < R_{\sf crit}$, we do not have a complete characterization. Let us consider the following choice of the auxiliary random variable $V$. We set 
\begin{IEEEeqnarray}{rCl} \label{eq: choice_V}
V &=& \begin{cases}
X^{(1)} \oplus U  & \text{ if } X^{(1)} = X^{(2)} , \\
W & \text{ if } X^{(1)}\neq X^{(2)}, 
\end{cases} 
\end{IEEEeqnarray}
where $\oplus$ denotes modulo-two sum, $U,W\in\{0,1\}$ are independent of $(X,Y)$, and furthermore $W\sim$ Bernoulli($1/2$). Numerical studies suggest that such choice of $V$ characterizes the boundary of $\mathcal{R}^\star$ for $R_{\sf c} < R_{\sf crit}$. It can be checked that setting 
\begin{IEEEeqnarray*}{rCCCl}
p_{U}(1) &=& \frac{1}{2} - \frac{\sqrt{1-2q}}{2(1-q)} &=:& \gamma
\end{IEEEeqnarray*}
achieves Wyner's common information $R_{\sf crit}$. 

In Figure \ref{fig:ex_single} we plot three inner bounds and an outer bound for the case $q=0.1$, where $R_{\sf crit}\approx 0.873$. The first inner bound is plotted in green dot, which results from memory sharing between the extreme points $(R_{\sf c}^\star,0)$ and $(0,R_{\sf u}^\star)$. The extreme point $\left(R_{\sf crit},\frac{1}{2}(1+h_{\sf b}(q)-R_{\sf crit})\right)$ is marked by a blue diamond point. Then, the second inner bound is formed by memory sharing among the three extreme points. The third inner bound has the same boundary as the second inner bound for $R_{\sf c} \ge R_{\sf crit}$. As for $R_{\sf c} < R_{\sf crit}$, the third inner bound is plotted in red solid, which results from evaluating all $p_{U}(1)\in[\gamma,0.5]$. Finally, the combined outer bound $R_{\sf u} \ge \max\{\frac{1}{2}(1+h_{\sf b}(q)-R_{\sf c})^+,(1-R_{\sf c})^+ \}$ is plotted in black solid.  

We remark that it can be shown that $H(X^{(n)}\oplus V) = H(X^{(n)}|V)$, $n\in\{1,2\}$. Thus, instead of Slepian--Wolf coding, the update encoder can simply compress $X^{(n)}\oplus V$ and transmit. Then, after recovering $X^{(n)}\oplus V$, the decoder removes $V$ to get the desired component $X^{(n)}$.
\hfill$\lozenge$\end{example}

\section{The Two-User Extensions} \label{sec:two_user}

\subsection{Sequential Coding over the Gray--Wyner Systems} \label{subsec:GW}
\begin{figure}[t!]
\begin{center}
\includegraphics[scale=0.75]{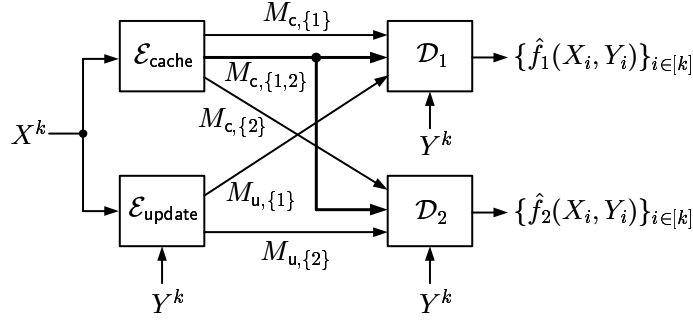}
\end{center}
\vspace{-0.25in}
\caption{The private-update-aided Gray--Wyner system.}
\label{fig:GW1}
\vspace{-0.1in}
\end{figure}

\begin{figure}[t!]
\begin{center}
\includegraphics[scale=0.75]{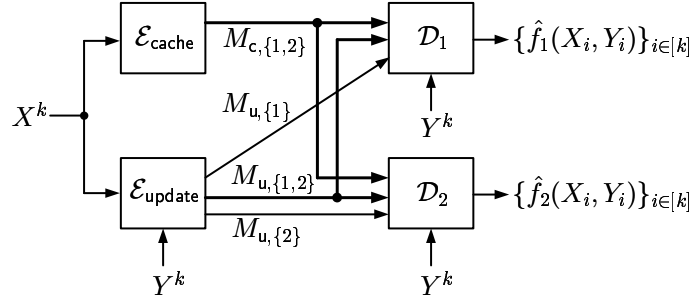}
\end{center}
\vspace{-0.25in}
\caption{The common-cache-aided Gray--Wyner system.}
\label{fig:GW2}
\vspace{-0.1in}
\end{figure}

Figures \ref{fig:GW1} and \ref{fig:GW2} depict the private-update-aided Gray--Wyner system and the common-cache-aided Gray--Wyner system, respectively. In the private-update-aided Gray--Wyner system, the cache encoder plays the role of encoder in the Gray--Wyner system and is aided by the update encoder through the private updates. By contrast, in the common-cache-aided Gray--Wyner system, the update encoder plays the role of encoder in the Gray--Wyner system and is aided by the cache encoder through the common cache. By applying sequential coding over the Gray--Wyner systems, we have the single-letter characterization for the considered setups. The proofs follow similar lines as in the single-user case and thus are omitted.

\begin{theorem}\label{thm:u2c1}
Consider the private-update-aided Gray--Wyner system. The optimal rate region $\mathcal{R}_{\sf puGW}^\star$ is the set of rate tuples $\mathbf{R}$ such that 
\begin{IEEEeqnarray*}{rCl}
R_{{\sf c},\{1,2\}} &\ge& I(X;V_{\sf c}|Y), \\
R_{{\sf c},\{1\}} &\ge& I(X;V_1|V_{\sf c},Y), \\
R_{{\sf c},\{2\}} &\ge& I(X;V_2|V_{\sf c},Y), \\
R_{{\sf u},\{1\}} &\ge& H(f_1(X,Y)|V_{\sf c},V_1,Y), \\
R_{{\sf u},\{2\}} &\ge& H(f_2(X,Y)|V_{\sf c},V_2,Y), 
\end{IEEEeqnarray*}
for some conditional pmf $p_{V_{\sf c}|X}p_{V_1|V_{\sf c},X}p_{V_2|V_{\sf c},X}$ satisfying $|\mathcal{V}_{\sf c}|\le |\mathcal{X}|+4$, $|\mathcal{V}_j| \le |\mathcal{V}_{\sf c}||\mathcal{X}|+1$, $j\in\{1,2\}$.
\end{theorem}

\begin{theorem}\label{thm:u2c2}
Consider the common-cache-aided Gray--Wyner system. The optimal rate region $\mathcal{R}_{\sf ccGW}^\star$ is the set of rate tuples $\mathbf{R}$ such that 
\begin{IEEEeqnarray*}{rCl}
R_{{\sf c},\{1,2\}} &\ge& I(X;V_{\sf c}|Y), \\
R_{{\sf u},\{1,2\}} &\ge& I(X;V_{\sf u}|V_{\sf c},Y), \\
R_{{\sf u},\{1\}} &\ge& H(f_1(X,Y)|V_{\sf c},V_{\sf u},Y), \\
R_{{\sf u},\{2\}} &\ge& H(f_2(X,Y)|V_{\sf c},V_{\sf u},Y), 
\end{IEEEeqnarray*}
for some conditional pmf $p_{V_{\sf c}|X}p_{V_{\sf u}|V_{\sf c},X,Y}$ satisfying $|\mathcal{V}_{\sf c}|\le |\mathcal{X}|+3$, $|\mathcal{V}_{\sf u}| \le |\mathcal{V}_{\sf c}||\mathcal{X}||\mathcal{Y}|+2$.
\end{theorem}

\subsection{Sequential Successive Refinement} \label{subsec:SeqSR}
\begin{figure}[t!]
\begin{center}
\includegraphics[scale=0.75]{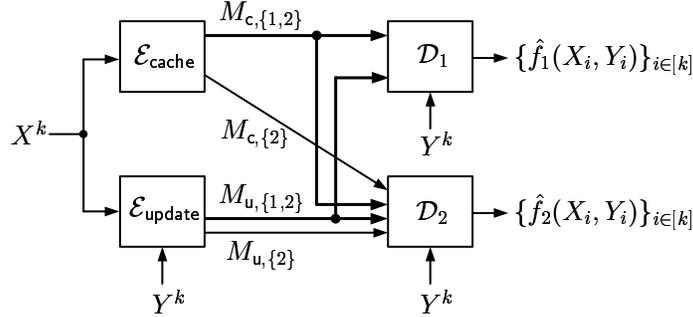}
\end{center}
\vspace{-0.25in}
\caption{The problem of sequential successive refinement.}
\label{fig:SR}
\vspace{-0.1in}
\end{figure}

Figure \ref{fig:SR} plots the problem of sequential successive refinement, which can be seen as a special case of the distributed successive refinement problem. In the first stage, the cache encoder and the private encoder each send a coarse description to both decoders, and then in the second stage they each send a refined description only to Decoder $2$. For this setup, we have a single-letter characterization for the optimal rate region $\mathcal{R}_{\sf SSR}^\star$ as given in the following theorem.

\begin{theorem}\label{thm:u2c3}
Consider the problem of sequential successive refinement. The optimal rate region $\mathcal{R}_{\sf SSR}^\star$ is the set of rate tuples $\mathbf{R}$ such that 
\begin{IEEEeqnarray*}{rCl}
R_{{\sf c},\{1,2\}} &\ge& I(X;V_{\sf c}|Y), \\
R_{{\sf c},\{1,2\}} + R_{{\sf c},\{2\}} &\ge&  I(X;V_{\sf c}|Y) + I(X;V_2|f_1(X,Y),V_{\sf c},Y), \\
R_{{\sf u},\{1,2\}} &\ge& H(f_1(X,Y)|V_{\sf c},Y), \\
R_{{\sf u},\{1,2\}} + R_{{\sf u},\{2\}} &\ge& H(f_1(X,Y)|V_{\sf c},Y) + H(f_2(X,Y)|V_2,f_1(X,Y),V_{\sf c},Y),
\end{IEEEeqnarray*}
for some conditional pmf $p_{V_2,V_{\sf c}|X}$ satisfying $|\mathcal{V}_{\sf c}|\le |\mathcal{X}|+3$ and $|\mathcal{V}_2|\le |\mathcal{V}_{\sf c}||\mathcal{X}|+1$.
\end{theorem}

\begin{IEEEproof}
The converse part is deferred to Appendix \ref{sec:conv_SSR}. The achievability can be proved by applying Theorem \ref{thm:single} and its straightforward extension. Here we provide a high-level description. Consider a simple two-stage coding. In the first stage, we use a multiple description code which follows the achievability for the single-user case and each encoder sends its generated message through its common link. Since both messages $(M_{{\sf c},\{1,2\}},M_{{\sf c},\{2\}})$ are also received by Decoder $2$, Decoder $2$ can also learn $(v_{\sf c}^k,\{f_1(x_i,y_i)\}_{i\in[k]})$. Then, in the second stage, we use another multiple description code which follows the achievability for the single-user case but with the augmented side information $(v_{\sf c}^k,\{f_1(x_i,y_i)\}_{i\in[k]},y^k)$. Once the messages are generated, each encoder can divide its message of the second stage into two parts, one of which is sent through the common link and the other is sent through the private link.
\end{IEEEproof}

\section{The Static Request Model} \label{sec:static}
Consider a database modeled by a DMS $\langle X\rangle$, which generates an i.i.d. source sequence $X^{Fk}$, where $k$ and $F$ are positive integers. The request is modeled by a DMS $\langle Y\rangle$, which generates an i.i.d. sequence $Y^{k}$. Here we assume that the request sequence $Y^k$ is independent of the database $X^{Fk}$. For all $i\in[k]$, we say that the $i$-th {\em block} consists of the subsequence $(X_{(i-1)F+1},X_{(i-1)F+2},\cdots,X_{iF})$ and the request $Y_i$. We assume that the user is interested in recovering $f(X_{(i-1)F+j},Y_i)$ for all $i\in[k]$ and $j\in[F]$. Then, the information-theoretic model corresponds to the case where $F=1$ and thus each $X_i$ is paired with a distinct $Y_i$, $i\in[k]$. Alternatively, one can think of it as processing 
\begin{IEEEeqnarray*} {rCl}
X_j,X_{F+j},X_{2F+j},X_{3F+j},\cdots ,X_{(k-1)F+j}
\end{IEEEeqnarray*}
for different $j\in[F]$ separately. We will refer to it as {\em coding across blocks}.

\begin{figure}[t!]
\begin{center}
\includegraphics[scale=0.7]{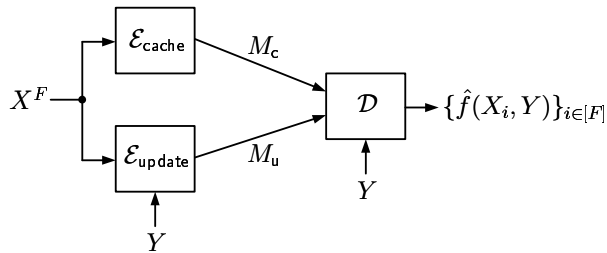}
\end{center}
\vspace{-0.15in}
\caption{The single-user case of the static request model.}
\label{fig:static request}
\vspace{-0.1in}
\end{figure}

In this section we consider the {\em static request model}, which corresponds to the case where $k=1$ (see Figure \ref{fig:static request}). Alternatively, one can think of it as processing 
\begin{IEEEeqnarray*} {rCl}
X_{(i-1)F+1},X_{(i-1)F+2},\cdots,X_{(i-1)F+F} 
\end{IEEEeqnarray*}
for different $i\in[k]$ separately. We will refer to it as {\em coding within block}. We remark that both coding across blocks and coding within block are special cases of {\em coding over multiple instances}, i.e., coding over the whole source sequence $X^{Fk}$.

We can draw an analogy between the request models for content delivery networks and the fading models for wireless networks. If we think of the requests as channel states, then the requests behave like {\em fast fading} in the information-theoretic model. On the other hand, in the static request model, the requests behave like {\em quasi-static fading}. 

For convenience, we assume that $\mathcal{Y}=[N]$, where $N$ is a positive integer. In the static request model, the cache encoder observes the source sequence $X^F$, the update encoder observes the source sequence $X^F$ and the single request $Y_1$, and the decoder only observes the request $Y_1$.\footnote{For notational convenience, we drop the subscript $1$ in $Y_1$ hereafter.} The cache encoder generates a message $M_{\sf c}\in[2^{FR_{\sf c}}]$, and the update encoder generates a message $M_{\sf u}(Y)\in[2^{FR_{\sf u}(Y)}]$. The decoder receives $(M_{\sf c},M_{\sf u}(Y))$ and wishes to recover the sequence of functions $\{f(X_j,Y)\}_{j\in[F]}$ losslessly. Note that the length of the update message is generally a function of $Y$.

A $(2^{FR_{\sf c}},\{2^{FR_{\sf u}(y)}\}_{y\in[N]},F)$ distributed multiple description code consists of 
\begin{itemize}
\item one cache encoder, which assigns an index $m_{\sf c}(x^F)\in[2^{FR_{\sf c}}]$ to each sequence $x^F\in\mathcal{X}^F$;
\item one update encoder, which assigns $N$ indices $m_{\sf u}(x^F,y)\in[2^{FR_{\sf u}(y)}]$, $y\in[N]$, to each sequence $x^F\in\mathcal{X}^F$;  
\item one decoder, which assigns an estimate $\hat{s}^F$ to each tuple $(m_{\sf c}, m_{\sf u}, y)$.
\end{itemize}
In words, there are $N$ codebooks at the update encoder and the decoder, each of which is served for one request $y\in[N]$. On the other hand, since the cache encoder does not observe the request $Y$ and cannot infer anything about $Y$ from the observed source sequence $X^F$, the cache encoder only needs one codebook. 

For the static request model, a rate tuple $(R_{\sf c},\{R_{\sf u}(y)\}_{y\in[N]})$ is said to be achievable if there exists a sequence of $(2^{FR_{\sf c}},\{2^{FR_{\sf u}(y)}\}_{y\in[N]},F)$ codes such that
\begin{IEEEeqnarray}{rCl} \label{eq:Pe_ally}
\lim_{F\to\infty} \mathbb{P}\left(\bigcup_{j\in[F]}\left\{\hat{S}_j\neq f(X_j,y)\right\}\right) &=& 0, \quad \forall y\in[N]. 
\end{IEEEeqnarray}
The optimal rate region $\mathcal{R}_{\sf static}^\star$ is the closure of the set of achievable rate pairs.

The above problem can be considered as an $N$-user Gray--Wyner system: Each user $y\in[N]$ receives a private message $M_{\sf u}(y)$ and a common message $M_{\sf c}$. The common message is received by all users. Each user $y\in [N]$ wishes to recover the sequence $\{f(X_j,y)\}_{j\in[F]}$. Then, we have the following theorem. 
\begin{theorem} \label{thm:singleY}
Consider the single-user case under the static request model. The optimal rate region $\mathcal{R}_{\sf static}^\star$ is the set of rate tuples $(R_{\sf c},\{R_{\sf u}(y)\}_{y\in[N]})$ such that 
\begin{IEEEeqnarray*}{rCl}
R_{\sf c} &\ge& I(X;V),  \\
R_{\sf u}(y) &\ge& H(f(X,y)|V), \quad \forall y\in[N], 
\end{IEEEeqnarray*} 
for some conditional pmf $p_{V|X}$ with $|\mathcal{V}| \le |\mathcal{X}|+N$.
\end{theorem}

The static request model can be studied under more specific performance criteria, which relate the request-dependent update rates $\{R_{\sf u}(y)\}$ to a fixed quantity $R_{\sf u}$, which is independent of the realization of $Y$. In the following, we discuss two common performance criteria by analogy with quasi-static fading. For convenience, we denote 
\begin{IEEEeqnarray*}{rCl}
\mathcal{R}^\star(R_{\sf c}) &=& \{\mathbf{R}: (r,\mathbf{R})\in\mathcal{R}_{\sf static}^\star, r\le R_{\sf c}\}, 
\end{IEEEeqnarray*}
where $\mathbf{R} = (R_{\sf u}(1),R_{\sf u}(2),\cdots,R_{\sf u}(N))$.

\subsection{Compound (The Worst Case)}
Consider a fixed update rate $R_{\sf u}$. The compound formulation requires that for each block of length $F$, the update message cannot contain more than $\lfloor FR_{\sf u}\rfloor$ bits. Namely, it requires that for all $y\in[N]$, 
\begin{IEEEeqnarray*}{rCl}
R_{\sf u}(y) &\le& R_{\sf u}.
\end{IEEEeqnarray*}
We remark that this setup is equivalent to the worst-case formulation considered by Maddah-Ali and Niesen in \cite{Maddah-Ali:14}.
Thus, the compound-optimal update rate given the cache rate $R_{\sf c}$ can be defined as 
\begin{IEEEeqnarray*}{rCl}
R_{\sf compound}(R_{\sf c}) &:=& \min_{\mathbf{R} \in \mathcal{R}^\star(R_{\sf c})} \max_{y\in[N]} R_{\sf u}(y). 
\end{IEEEeqnarray*}
Then, from Theorem \ref{thm:singleY}, we have the following corollary.
\begin{corollary}
Consider the compound formulation of the static request model. The compound-optimal update rate given the cache rate $R_{\sf c}$ can be expressed as 
\begin{IEEEeqnarray*}{rCl}
R_{\sf compound}(R_{\sf c}) &=& \min_{\substack{ p_{V|X} \\  \text{s.t. } I(X;V)\le R_{\sf c}}} \max_{y\in[N]}H(f(X,y)|V), 
\end{IEEEeqnarray*}
where $|\mathcal{V}| \le |\mathcal{X}|+N$. 
\end{corollary}

The compound formulation aims to model the worst-case scenario in which the request statistics is not known and/or the communication resource cannot be redistributed over blocks. This formulation is also studied in \cite{Timo:16}.
Finally, we remark that one may relax the compound formulation by tolerating some {\em outage} events, which requires higher update rates than a predefined value. More details of the outage formulation can be found in \cite[Chapter 3.7.2]{CYWang:15}.

\subsection{Adaptive Coding (The Average Case)}
Different from the compound formulation, here the unused communication resource can be saved for the other blocks. The only requirement is that the average number of bits per block cannot be larger than $\lfloor FR_{\sf u}\rfloor$. This setup is equivalent to the average-case formulation considered by Niesen and Maddah-Ali in \cite{Niesen:14}. To see the relation of the adaptive coding formulation with the information-theoretic model, let us consider the whole rate region instead. We define the adaptive-optimal rate region as 
\begin{IEEEeqnarray*}{rCl}
\mathcal{R}_{\sf adaptive} &:=& \left\{\left(R_{\sf c},\mathbb{E}_Y[R_{\sf u}(Y)]\right): (R_{\sf c},\{R_{\sf u}(y)\}_{y\in[N]})\in\mathcal{R}^\star\right\}.
\end{IEEEeqnarray*}
From Theorem \ref{thm:singleY} we have the following corollary. 
\begin{corollary} \label{col:adaptive}
Consider the adaptive coding formulation of the static request model. The adaptive-optimal rate region $\mathcal{R}_{\sf adaptive}$ is the set of rate pairs $(R_{\sf c},R_{\sf u})$ such that 
\begin{IEEEeqnarray}{rCl}
\label{eq:single_Yc}
R_{\sf c} &\ge& I(X;V),  \\
\label{eq:single_Yu}
R_{\sf u} &\ge& H(f(X,Y)|V,Y),
\end{IEEEeqnarray} 
for some conditional pmf $p_{V|X}$ with $|\mathcal{V}| \le |\mathcal{X}|+1$.
\end{corollary}
\begin{IEEEproof}
It suffices to show that $\mathbb{E}_Y[\phi(Y)] = H(f(X,Y)|V,Y)$, where $\phi(y) = H(f(X,y)|V)$. Indeed, we have 
\begin{IEEEeqnarray*}{rCl}
E_Y[\phi(Y)] &=& \sum_{y=1}^N p_Y(y) \phi(y) \\
&=& \sum_{y=1}^N p_Y(y)H(f(X,y)|V) \\
&\overset{(a)}{=}& \sum_{y=1}^N p_Y(y)H(f(X,y)|V,Y=y) \\
&=& H(f(X,Y)|V,Y), 
\end{IEEEeqnarray*} 
where $(a)$ follows since $X$ is independent of $Y$, by assumption, and $V\markov X\markov Y$ form a Markov chain. Finally, we remark that the cardinality bound on $\mathcal{V}$ is refined from $|\mathcal{X}|+N$ to $|\mathcal{X}|+1$, which can be proved using the convex cover method \cite[Appendix C]{ElGamal:11}. 
\end{IEEEproof}

For the single-user caching problem, Corollary \ref{col:adaptive} thus shows that the
rate region for the information-theoretic model (Theorem \ref{thm:single}) takes exactly the same shape as the rate region for the average-case of the Maddah-Ali--Niesen model (Equations \eqref{eq:single_Yc} and \eqref{eq:single_Yu}) in spite of the fact that the modeling assumptions are rather different. Beyond the single-user case, this equivalence of rate regions continues to hold at least for the extended Gray--Wyner systems studied in Section \ref{subsec:GW}, which can be established by arguments along the lines of the proof of Corollary \ref{col:adaptive}. In more general multi-user cases, however, the rate region of the information-theoretic model is {\em different} from the rate region of the Maddah-Ali--Niesen model, as the following example illustrates:

\begin{figure}[t!]
\begin{center}
\includegraphics[scale=0.7]{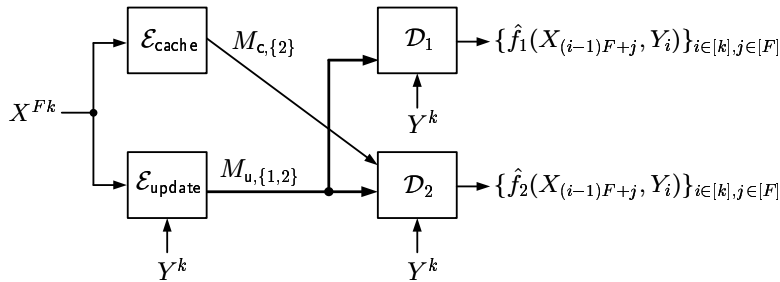}
\end{center}
\vspace{-0.15in}
\caption{The system considered in Example \ref{ex:Fk}. The information-theoretic model corresponds to the case where $F=1$ and the static request model corresponds to the case where $k=1$.}
\label{fig:sys_asym}
\vspace{-0.1in}
\end{figure}

\begin{example} \label{ex:Fk}
Consider the system depicted in Figure \ref{fig:sys_asym}. The system is a special case of sequential successive refinement, discussed in Section \ref{subsec:SeqSR}, in which $R_{{\sf c},\{1,2\}}=R_{{\sf u},\{2\}}=0$. We assume that $X\sim$ Bernoulli($1/2$), $Y\sim$ Uniform($\{1,2\}$), $f_1(X,Y)=X\cdot\mathbbm{1}\{Y=1\}$, and $f_2(X,Y)=X$, where $\mathbbm{1}\{\cdot\}$ is the indicator function. We observe that if $Y=1$, then the cache content at User $2$ is useless because the update encoder has to send $X$ to User $1$ anyway. 

Following from Theorem \ref{thm:u2c3}, the optimal rate region under the information-theoretic model, denoted by $\mathcal{R}_{\sf it}^\star$, is the the set of rate pairs $(R_{{\sf c},\{2\}},R_{{\sf u},\{1,2\}})$ such that 
\begin{IEEEeqnarray*}{rCl}
R_{{\sf c},\{2\}} &\ge&  I(X;V|f_1(X,Y),Y), \\
R_{{\sf u},\{1,2\}} &\ge& H(f_1(X,Y)|Y) + H(f_2(X,Y)|V,f_1(X,Y),Y),
\end{IEEEeqnarray*}
for some conditional pmf $p_{V|X}$. After specializing to the considered instance, we have the following closed-form expression:
\begin{IEEEeqnarray*}{rCl}
R_{{\sf c},\{2\}} &\ge& \frac{1}{2}(0+r) = \frac{r}{2}, \\
R_{{\sf u},\{1,2\}} &\ge& \frac{1}{2} + \frac{1}{2}(1-r)^+,
\end{IEEEeqnarray*}
for some $r\ge0$.

On the other hand, although a single-letter characterization for the optimal rate region under the static request model, denoted by $\mathcal{R}_{\sf sr2}^\star$, is unknown in general, we have a closed-form expression for the considered instance: $\mathcal{R}_{\sf sr2}^\star$ is the set of rate tuples $(R_{{\sf c},\{2\}},R_{{\sf u},\{1,2\}}(1),R_{{\sf u},\{1,2\}}(2))$ such that 
\begin{IEEEeqnarray*}{rCl}
R_{{\sf c},\{2\}} &\ge& \max\{0,r\} = r, \\
R_{{\sf u},\{1,2\}}(1) &\ge& 1, \\
R_{{\sf u},\{1,2\}}(2) &\ge& (1-r)^+, 
\end{IEEEeqnarray*}
for some $r\ge 0$. 

For the achievability, we borrow the coding insights gained from the information-theoretic model, in which Decoder $2$ decodes $f_1(X,Y)$ first and uses it as side information. For the static request model all rate tuples $(R_{{\sf c},\{2\}},R_{{\sf u},\{1,2\}}(1),R_{{\sf u},\{1,2\}}(2))$ satisfying 
\begin{IEEEeqnarray*}{rCl}
R_{{\sf c},\{2\}} &>& \max_{y\in\{1,2\}} I(X;V|f_1(X,y)), \\
R_{{\sf u},\{1,2\}}(y) &>& H(f_1(X,y)) + H(f_2(X,y)|V,f_1(X,y)), \quad y\in \{1,2\},
\end{IEEEeqnarray*}
for some conditional pmf $p_{V|X}$ are achievable. Then, the achievability follows by setting $V=(XQ,Q)$, where $Q\sim$ Bernoulli($\min\{r,1\}$) is independent of $(X,Y)$. 

To see the optimality, we note that if $Y=1$, then Decoder $1$ must be able to recover $X^F$ losslessly from $M_{{\sf u},\{1,2\}}(1)$ and thus we have $R_{{\sf u},\{1,2\}}(1) \ge 1$. If $Y=2$, then Decoder $2$ must be able to recover $X^F$ losslessly from $M_{{\sf c},\{2\}}$ and $M_{{\sf u},\{1,2\}}(2)$, which gives the sum rate constraint $R_{{\sf c},\{2\}}+R_{{\sf u},\{1,2\}}(2) \ge 1$.

Therefore, the corresponding adaptive-optimal rate region is the set of rate pairs 
$(R_{{\sf c},\{2\}},R_{{\sf u},\{1,2\}})$ such that 
\begin{IEEEeqnarray*}{rCl}
R_{{\sf c},\{2\}} &\ge& r, \\
R_{{\sf u},\{1,2\}} &\ge& \frac{1}{2} + \frac{1}{2}(1-r)^+,
\end{IEEEeqnarray*}
for some $r\ge0$. 
\hfill$\lozenge$\end{example}

The above example shows that coding across blocks can be more beneficial than coding within block and we conjecture that it is always the case. Furthermore, we conjecture that whenever coding across blocks is feasible, coding within block provides no additional gain. 

\section{Conclusion} \label{sec:conclude}
In this paper, we have formulated the caching problem as a multi-terminal source coding problem with side information. The key observation is that we can treat the requested data as a function of the whole data and the request. All forms of data and requests can be simply modeled by random variables $X$ and $Y$, respectively, and their relation can be simply described by a function $f$. Thanks to the formulation, many coding techniques and insights can be directly borrowed from the well-developed source coding literature. For the single-user case, we have given a single-letter characterization of the optimal rate region and found closed-form expressions for two interesting cases. Then, using the insights gained from the Gray--Wyner system and the successive refinement problem, we were able to find single-letter expressions of optimal rate region for three two-user scenarios. Finally, we showed that although the information-theoretic model and the static request model have a similar behavior for the single-user case, they in general behave differently in multi-user scenarios. 

\appendices
 
\section{Proof of Theorem \ref{thm:single}} \label{sec:proof_single}
{\em (Achievability.)} The proof follows from standard random coding arguments as in the problem of lossless source coding with a helper. Here we provide a high-level description of the coding scheme. First, the cache encoder applies Wyner--Ziv coding on $x^k$ given side information $y^k$ so that the decoder learns $v^k$, a quantized version of $x^k$. Then the update encoder applies Slepian--Wolf coding on the function sequence $(f(x_i,y_i):i\in[k])$ given side information $(v^k,y^k)$.

{\em (Converse.)} Denote $S_i = f(X_i,Y_i)$, $i\in[k]$. First, we have 
\begin{IEEEeqnarray*}{rCl}
kR_{\sf c} &\ge & H(M_{\sf c}) \\
&\ge& H(M_{\sf c}|Y^k) \\
&= & I(X^k;M_{\sf c}|Y^k) \\
&=& \sum_{i=1}^k  I(X_i;M_{\sf c}|X^{i-1},Y^k) \\
&\overset{(a)}{=}& \sum_{i=1}^k  I(X_i;M_{\sf c},X^{i-1},Y^{[k]\backslash\{i\}}|Y_i) \\
&\ge& \sum_{i=1}^k I(X_i;V_i|Y_i), 
\end{IEEEeqnarray*}
where $(a)$ follows since $(X_i,Y_i)$ is independent of $(X^{i-1},Y^{[k]\backslash\{i\}})$. For the last step, we define $V_i := (M_{\sf c},S^{i-1},Y^{[k]\backslash\{i\}})$. Note that $V_i\markov X_i \markov Y_i$ form a Markov chain. 

Next, we have  
\begin{IEEEeqnarray*}{rCl}
 kR_{\sf u} &\ge& H(M_{\sf u}) \\
&\ge& H(M_{\sf u}|M_{\sf c},Y^k) \\
&=&  H(S^k,M_{\sf u}|M_{\sf c},Y^k) - H(S^k|M_{\sf c},M_{\sf u},Y^k) \\
&\overset{(a)}{\ge}& H(S^k|M_{\sf c},Y^k) -k\epsilon_k \\
&=& \sum_{i=1}^k H(S_i|M_{\sf c},S^{i-1}, Y^k) -k\epsilon_k \\
&=& \sum_{i=1}^k H(S_i|V_i, Y_i) -k\epsilon_k, 
\end{IEEEeqnarray*}
where $(a)$ follows from the data processing inequality and Fano's inequality. 
The rest of the proof follows from the standard time-sharing argument and then letting $k\to\infty$. The cardinality bound on $\mathcal{V}$ can be proved using the convex cover method \cite[Appendix C]{ElGamal:11}. 

\begin{remark}
First, the converse can also be established by identifying the auxiliary random variable $V_i = (M_{\sf c},X^{i-1},Y^{[k]\backslash\{i\}})$. Second, the optimal rate region $\mathcal{R}^\star$ is convex since the auxiliary random variable $V$ performs memory sharing implicitly. Finally, as can be seen from the achievability, even if the update encoder is restricted to access only the function sequence $(f(x_i,y_i):i\in[k])$, instead of $(x^k,y^k)$, the rate region remains the same. 
\hfill$\lozenge$\end{remark}

\section{Proof of Corollary \ref{col:single}} \label{sec:proof_property}
Statements $1$ and $2$ are straightforward. Statement $3$ can be proved by a simple cut-set argument. Alternatively, from Theorem \ref{thm:single} we observe that  
\begin{IEEEeqnarray*}{rCl}
R_{\sf c}+R_{\sf u} &\ge& I(X;V|Y) + H(f(X,Y)|V,Y) \\
&\ge& I(f(X,Y);V|Y) + H(f(X,Y)|V,Y) \\
&=& H(f(X,Y)|Y).
\end{IEEEeqnarray*}
Then, the lower bound on sum rate implies $R_{\sf c}^\star \ge H(f(X,Y)|Y) = R_{\sf u}^\star$, i.e., Statement $4$. 

Now we prove Statement $5$. Assume that $(R_{\sf c},R_{\sf u})\in \mathcal{R}^\star$ and fix any $a\in[0,R_{\sf c}]$. The case where $R_{\sf c}=0$ is trivial. Next, we consider the case where $R_{\sf c}>0$. Time sharing between $(0,R_{\sf u}^\star)$ and $(R_{\sf c},R_{\sf u})$ asserts that 
\begin{IEEEeqnarray*}{rCl}
\left(R_{\sf c}-a, R_{\sf u} + \frac{R_{\sf u}^\star-R_{\sf u}}{R_{\sf c}}a\right) \in \mathcal{R}^\star. 
\end{IEEEeqnarray*}
Then, Statements $1$ and $3$ imply that 
\begin{IEEEeqnarray*}{rCl}
\frac{R_{\sf u}^\star-R_{\sf u}}{R_{\sf c}} &\le& 1, 
\end{IEEEeqnarray*}
so it holds that $(R_{\sf c}-a,R_{\sf u}+a)\in \mathcal{R}^\star$.

Finally, we prove Statement $6$. Assume that $(R_{\sf c}+a,R_{\sf u}-a)\in \mathcal{R}^\star$. Then, memory sharing between $(0,R_{\sf u}^\star)$ and $(R_{\sf c}+a,R_{\sf u}-a)$ asserts that 
\begin{IEEEeqnarray*}{rCl}
\left(R_{\sf c}, \frac{aR_{\sf u}^\star+R_{\sf c}(R_{\sf u}-a)}{R_{\sf c}+a}\right) \in \mathcal{R}^\star. 
\end{IEEEeqnarray*}
However, Statement $3$ implies that 
\begin{IEEEeqnarray*}{rCl}
\frac{aR_{\sf u}^\star+R_{\sf c}(R_{\sf u}-a)}{R_{\sf c}+a} &\le& R_{\sf u}, 
\end{IEEEeqnarray*}
which contradicts the assumption that $(R_{\sf c},R_{\sf u})$ is an extreme point.

\section{Proof of Proposition \ref{prop:ind} (Converse)} \label{sec:conv_ind}
Suppose that $(R_{\sf c},R_{\sf u})\in\mathcal{R}^\star$. Then, there exists a conditional pmf $p_{V|X}$ such that $R_{\sf c} \ge I(X;V|Y) =: r$ and $R_{\sf u} \ge H(f(X,Y)|V,Y)$. For all $n\in[N]$, we have 
\begin{IEEEeqnarray*}{rCl}
r &=& I(X;V|Y) \\
& \overset{(a)}{=} & I(X;V) \\
&\ge& I(X^{(1)},\cdots,X^{(n)};V) \\
\label{eq:cond_ind}
&\ge& \sum_{j=1}^n H(X^{(j)}) - \sum_{j=1}^n H(X^{(j)}|V), \IEEEyesnumber
\end{IEEEeqnarray*}
where $(a)$ follows since $X$ and $Y$ are independent. Next we show that $R_{\sf u}$ can be lower bounded as in \eqref{eq:ind2}:
\begin{IEEEeqnarray*}{rCl}
R_{\sf u} &\ge& H(f(X,Y)|V,Y) \\
&=& \sum_{j=1}^N p_Y(j) H(X^{(j)}|V) \\
&\overset{(a)}{\ge}& p_Y(N)\left(\sum_{j=1}^N H(X^{(j)})- r - \sum_{j=1}^{N-1} H(X^{(j)}|V)\right)^+ + \sum_{j=1}^{N-1} p_Y(j) H(X^{(j)}|V) \\
\label{eq:ind_tmp}
&\overset{(b)}{\ge}& p_Y(N)\left(\sum_{j=1}^N H(X^{(j)})- r\right)^+ + \sum_{j=1}^{N-1} (p_Y(j)-p_Y(N)) H(X^{(j)}|V), \IEEEyesnumber
\end{IEEEeqnarray*}
where $(a)$ follows from \eqref{eq:cond_ind} with $n=N$ and $H(X^{(N)}|V) \ge 0$ and $(b)$ follows since $(u-v)^+\ge(u)^+-v$ for all $v\ge 0$.
The term on the right-hand side of \eqref{eq:ind_tmp} can be lower bounded as 
\begin{IEEEeqnarray*}{ll}
& \sum_{j=1}^{N-1} (p_Y(j)-p_Y(N)) H(X^{(j)}|V)  \\
&\overset{(a)}{\ge} (p_Y(N-1)-p_Y(N)) \left(\sum_{j=1}^{N-1} H(X^{(j)})- r - \sum_{j=1}^{N-2} H(X^{(j)}|V)\right)^+ \nonumber \\
&\hspace{0.5cm} + \sum_{j=1}^{N-2} (p_Y(j)-p_Y(N)) H(X^{(j)}|V) \\
&\ge (p_Y(N-1)-p_Y(N)) \left(\sum_{j=1}^{N-1} H(X^{(j)})- r\right)^+ + \sum_{j=1}^{N-2} (p_Y(j)-p_Y(N-1)) H(X^{(j)}|V), \IEEEeqnarraynumspace 
\end{IEEEeqnarray*}
where $(a)$ follows from \eqref{eq:cond_ind} with $n=N-1$ and $H(X^{(N-1)}|V) \ge 0$.
At this point, it is clear that we can apply the same argument for another $N-2$ times and arrive at 
\begin{IEEEeqnarray}{rCl}
\label{eq:lower_ind}
R_{\sf u} &\ge& \sum_{n=1}^N (p_Y(n)-p_Y(n+1)) \left(\sum_{j=1}^nH(X^{(j)})-r\right)^+, 
\end{IEEEeqnarray}
where $p_Y(N+1) = 0$. 

\section{Proof of Proposition \ref{prop:nest} (Converse)} \label{sec:conv_nest}
Suppose that $(R_{\sf c},R_{\sf u})\in\mathcal{R}^\star$. Then, there exists a conditional pmf $p_{V|X}$ such that $R_{\sf c} \ge I(X;V|Y) =: r$ and $R_{\sf u} \ge H(f(X,Y)|V,Y)$. For all $n\in[N]$, we have 
\begin{IEEEeqnarray*}{rCl}
r &=& I(X;V|Y) \\
& \overset{(a)}{=} & I(X;V) \\
&\ge& I(X^{(1)},\cdots,X^{(n)};V) \\
\label{eq:cond_nest}
&\overset{(b)}{=}& H(X^{(n)}) - \sum_{j=1}^n H(X^{(j)}|V,X^{(j-1)}), \IEEEyesnumber
\end{IEEEeqnarray*}
where $(a)$ follows since $X$ and $Y$ are independent and $(b)$ follows from the assumption that $H(X^{(n)}|X^{(n+1)}) = 0$ for all $n\in[N-1]$. Next, we show that $R_{\sf u}$ can be lower bounded as in \eqref{eq:nest2}: 
\begin{IEEEeqnarray*}{rCl}
R_{\sf u} &\ge& H(f(X,Y)|V,Y) \\
&=& \sum_{n=1}^N p_Y(n)H(X^{(n)}|V) \\
&\overset{(a)}{=}& \sum_{n=1}^N p_Y(n)H(X^{(1)},\cdots,X^{(n)}|V) \\
&\overset{(b)}{=}& \sum_{n=1}^N p_Y(n) \sum_{j=1}^n H(X^{(j)}|V,X^{(j-1)}) \\
&=& \sum_{j=1}^N \left(\sum_{n=j}^N p_Y(n)\right) H(X^{(j)}|V,X^{(j-1)}), 
\end{IEEEeqnarray*}
where $(a)$ and $(b)$ follow from the assumption that $H(X^{(n)}|X^{(n+1)}) = 0$ for all $n\in[N-1]$. For notational convenience, let us denote $s_j = \sum_{n=j}^N p_Y(n)$ and $q_j = H(X^{(j)}|V,X^{(j-1)})$. Then, we have 
\begin{IEEEeqnarray*}{rCl}
R_{\sf u} &\ge& \sum_{j=1}^N s_j q_j \\
&\overset{(a)}{\ge}& s_N\left(H(X^{(N)})-r-\sum_{j=1}^{N-1}q_j\right)^+ + \sum_{j=1}^{N-1} s_j q_j \\
&\overset{(b)}{\ge}& s_N\left(H(X^{(N)})-r\right)^+ + \sum_{j=1}^{N-1} (s_j-s_N) q_j \\
&=& p_Y(N)\left(H(X^{(N)})-r\right)^+ + \sum_{j=1}^{N-1} (s_j-s_N) q_j \\
&\overset{(c)}{\ge}& p_Y(N)\left(H(X^{(N)})-r\right)^+ + (s_{N-1}-s_N) \left(H(X^{(N-1)})-r-\sum_{j=1}^{N-2}q_j\right)^+ \nonumber\\
&& \hspace{0.2cm} + \sum_{j=1}^{N-2} (s_j-s_N) q_j \\
&\overset{(d)}{\ge}& p_Y(N)\left(H(X^{(N)})-r\right)^+ + (s_{N-1}-s_N) \left(H(X^{(N-1)})-r\right)^+  + \sum_{j=1}^{N-2} (s_j-s_{N-1}) q_j \IEEEeqnarraynumspace \\
&=& \sum_{n=N-1}^{N} p_Y(n)\left(H(X^{(n)})-r\right)^+ + \sum_{j=1}^{N-2} (s_j-s_{N-1}) q_j,
\end{IEEEeqnarray*}
where $(a)$ and $(c)$ follow from \eqref{eq:cond_nest} and $H(X^{(n)}|V,X^{(n-1)})\ge0$ with $n=N$ and $n=N-1$, respectively, and $(b)$ and $(d)$ follow since $(u-v)^+\ge(u)^+-v$ for all $v\ge 0$. At this point, it is clear that we can apply the same argument for another $N-2$ times and arrive at  
\begin{IEEEeqnarray}{rCl}
\label{eq:lower_nest}
R_{\sf u} &\ge& \sum_{n=1}^N p_Y(n)\left(H(X^{(n)})-r\right)^+. 
\end{IEEEeqnarray}

\section{Proof of Proposition \ref{prop:uni}} \label{sec:proof_uni}
Denote by $\overline{\mathcal{R}}$ the set of rate pairs $(R_{\sf c},R_{\sf u})$ such that 
\begin{IEEEeqnarray*}{rCl}
R_{\sf c} &\ge& I(\overline{X};V|Y), \\
R_{\sf u} &\ge& H(f(X,Y)|V,Y), 
\end{IEEEeqnarray*}
for some conditional pmf $p_{V|X}$. Since $I(X;V|Y)\ge I(\overline{X};V|Y)$, we have $\mathcal{R}^\star \subseteq \overline{\mathcal{R}}$. Also, it can be shown that the rate region $\overline{\mathcal{R}}$ is achievable, so we conclude that $\mathcal{R}^\star = \overline{\mathcal{R}}$. By using the assumptions that $X$ and $Y$ are independent and that $Y$ is uniformly distributed, we can simplify the rate expressions as 
\begin{IEEEeqnarray*}{rCl}
R_{\sf c} &\ge& I(\overline{X};V), \\
R_{\sf u} &\ge& \frac{1}{N} \sum_{n=1}^N H(X^{(n)}|V). 
\end{IEEEeqnarray*}

Now denote by $p_{V|\overline{X}}$ the conditional pmf induced by the conditional pmf $p_{V|X}$. As can be checked, both $I(\overline{X};V)$ and $\{H(X^{(n)}|V)\}_{n\in[N]}$ can be completely determined by the induced conditional pmf $p_{V|\overline{X}}$. Thus, 
it suffices to consider the space of all conditional pmfs $p_{V|\overline{X}}$. Finally, noting that 
\begin{IEEEeqnarray*}{rCl}
\sum_{n=1}^N H(X^{(n)}|V) &=& H(\overline{X}) - I(\overline{X};V) + \Gamma(\overline{X}|V), 
\end{IEEEeqnarray*}
it holds that if $R_{\sf c} = r \in [0,H(\overline{X})]$, then
\begin{IEEEeqnarray*}{rCl}
\min \{R_{\sf u} | R_{\sf c}=r, (R_{\sf c},R_{\sf u})\in\mathcal{R}^\star\} &=&  \frac{1}{N}\left(H(\overline{X})-r+\min_{\substack{ p_{V|\overline{X}} \\  \text{s.t. } I(\overline{X};V) = r}}\Gamma(\overline{X}|V)\right).
\end{IEEEeqnarray*}

\section{Proof of Theorem \ref{thm:u2c3} (Converse)} \label{sec:conv_SSR}
Denote $S_{1i}=f_1(X_i,Y_i)$ and $S_{2i}=f_2(X_i,Y_i)$ for $i\in[k]$. The rates $R_{{\sf c},\{1,2\}}$ and $R_{{\sf u},\{1,2\}}$ can be lower bounded in the same manner as the single-user case and thus the details are omitted. Denote $V_{{\sf c} i} = (M_{{\sf c},\{1,2\}},S_1^{i-1},Y^{[k]\backslash\{i\}})$. Now consider the bounds on $R_{{\sf c},\{1,2\}} +R_{{\sf c},\{2\}} $ and $R_{{\sf u},\{1,2\}} + R_{{\sf u},\{2\}}$. First, we have 
\begin{IEEEeqnarray*}{ll}
& k(R_{{\sf c},\{1,2\}} + R_{{\sf c},\{2\}}) \\
&\ge H(M_{{\sf c},\{1,2\}},M_{{\sf c},\{2\}}|Y^k) \\
&= I(X^k;M_{{\sf c},\{1,2\}},M_{{\sf c},\{2\}}|Y^k) \\
&= I(S_1^k,X^k;M_{{\sf c},\{1,2\}},M_{{\sf c},\{2\}}|Y^k) \\
&\ge I(S_1^k;M_{{\sf c},\{1,2\}}|Y^k) + I(X^k;M_{{\sf c},\{1,2\}},M_{{\sf c},\{2\}}|S_1^k,Y^k) \\
&= \sum_{i=1}^k I(S_{1i};M_{{\sf c},\{1,2\}}|S_1^{i-1},Y^k) + \sum_{i=1}^k I(X_i;M_{{\sf c},\{1,2\}},M_{{\sf c},\{2\}}|X^{i-1},S_1^k,Y^k) \\
&\overset{(a)}{=} \sum_{i=1}^k I(S_{1i};M_{{\sf c},\{1,2\}},S_1^{i-1},Y^{[k]\backslash\{i\}}|Y_i) \\
&\hspace{0.5cm}   + \sum_{i=1}^k I(X_i;M_{{\sf c},\{1,2\}},S_1^{i-1},Y^{[k]\backslash\{i\}},M_{{\sf c},\{2\}},X^{i-1},S_{1,i+1}^{k}|S_{1i},Y_i) \\
&= \sum_{i=1}^k I(S_{1i};V_{{\sf c}i}|Y_i) + \sum_{i=1}^k I(X_i;V_{{\sf c}i},V_{2i}|S_{1i},Y_i), 
\end{IEEEeqnarray*}
where $(a)$ follows since $(X_i,Y_i,S_{1i})$ is independent of $(X^{i-1},S_1^{[k]\backslash\{i\}},Y^{[k]\backslash\{i\}})$. For the last step, we define $V_{2i} := (M_{{\sf c},\{2\}},X^{i-1},S_{1,i+1}^k)$. Note that $(V_{{\sf c}i},V_{2i})\markov X_i \markov Y_i$ form a Markov chain. 
Second, we have 
\begin{IEEEeqnarray*}{ll}
& k(R_{{\sf u},\{1,2\}}+R_{{\sf u},\{2\}}) \\
&\ge H(M_{{\sf u},\{1,2\}}|M_{{\sf c},\{1,2\}},Y^k) + H(M_{{\sf u},\{2\}})\\
&\overset{(a)}{\ge}  H(S_1^k,M_{{\sf u},\{1,2\}}|M_{{\sf c},\{1,2\}},Y^k) + H(M_{{\sf u},\{2\}}) -k\epsilon_k' \\
&= \sum_{i=1}^k H(S_{1i}|M_{{\sf c},\{1,2\}},S_1^{i-1}, Y^k) + H(M_{{\sf u},\{1,2\}}|S_1^k,M_{{\sf c},\{1,2\}},Y^k) + H(M_{{\sf u},\{2\}}) -k\epsilon_k' \\
&\ge \sum_{i=1}^k H(S_{1i}|V_{{\sf c}i},Y_i) + H(M_{{\sf u},\{1,2\}},M_{{\sf u},\{2\}}|S_1^k,M_{{\sf c},\{1,2\}},M_{{\sf c},\{2\}},Y^k) -k\epsilon_k' \\
&\overset{(b)}{\ge} \sum_{i=1}^k H(S_{1i}|V_{{\sf c}i},Y_i) + H(S_2^k,M_{{\sf u},\{1,2\}},M_{{\sf u},\{2\}}|S_1^k,M_{{\sf c},\{1,2\}},M_{{\sf c},\{2\}},Y^k) -k(\epsilon_k'+\epsilon_k'') \IEEEeqnarraynumspace \\
&\ge \sum_{i=1}^k H(S_{1i}|V_{{\sf c}i},Y_i) + H(S_2^k|S_1^k,M_{{\sf c},\{1,2\}},M_{{\sf c},\{2\}},Y^k) -k(\epsilon_k'+\epsilon_k'') \\
&= \sum_{i=1}^k H(S_{1i}|V_{{\sf c}i},Y_i) + \sum_{i=1}^k H(S_{2i}|S_2^{i-1},S_1^k,M_{{\sf c},\{1,2\}},M_{{\sf c},\{2\}},Y^k) -k(\epsilon_k'+\epsilon_k'') \\
&\ge \sum_{i=1}^k H(S_{1i}|V_{{\sf c}i},Y_i) + \sum_{i=1}^k H(S_{2i}|S_{1i},V_{2i},V_{{\sf c}i},Y_i) -k(\epsilon_k'+\epsilon_k''), 
\end{IEEEeqnarray*}
where $(a)$ and $(b)$ follow from the data processing inequality and Fano's inequality.
The rest of the proof follows from the standard time-sharing argument, letting $k\to\infty$, and the fact that 
\begin{IEEEeqnarray*}{ll}
& I(f_1(X,Y);V_{\sf c}|Y) + I(X;V_{\sf c},V_2|f_1(X,Y),Y) \\
&= I(X;V_{\sf c}|Y) + I(X;V_2|f_1(X,Y),V_{\sf c},Y).
\end{IEEEeqnarray*}
The cardinality bounds on $\mathcal{V}_{\sf c}$ and $\mathcal{V}_2$ can be proved using the convex cover method \cite[Appendix C]{ElGamal:11}. 

\bibliographystyle{IEEEtran}
\bibliography{IEEEabrv,References}
\end{document}